\documentclass[float,aps,prc,superscriptaddress,showpacs,twocolumn,longbibliography]{revtex4-1}
\usepackage{bm}
\usepackage{amsmath}
\usepackage{amssymb}
\usepackage{color}
\usepackage{epsfig}
\usepackage{hyperref}
\usepackage{MnSymbol}
\usepackage[titletoc]{appendix}
\usepackage{simplewick}
\usepackage{multirow}
\usepackage[normalem]{ulem}
\usepackage[bottom]{footmisc}
\newcommand{\be}{\begin{equation}}
\newcommand{\ee}{\end{equation}}

\newcommand{\bes}{\begin{split}}
\newcommand{\ees}{\end{split}}
\newcommand{\ber}{\begin{eqnarray}}
\newcommand{\eer}{\end{eqnarray}}

\begin{document}

\title{Computing the dipole polarizability of $^{48}$Ca with increased precision}

\author{M. Miorelli} \affiliation{TRIUMF, 4004 Wesbrook Mall,
  Vancouver, British Columbia, V6T 2A3, Canada}
\affiliation{Department of Physics
  and Astronomy, University of British Columbia, Vancouver, British Columbia, V6T
  1Z4, Canada}

\author {S. Bacca}
\affiliation{Institut f\"ur Kernphysik and PRISMA Cluster of
  Excellence, Johannes Gutenberg-Universit\"at Mainz, 55128 Mainz,
  Germany}
\affiliation{TRIUMF, 4004 Wesbrook Mall, Vancouver, British Columbia,
  V6T 2A3, Canada}
\affiliation{Department of Physics and Astronomy, University of
  Manitoba, Winnipeg, Manitoba, R3T 2N2, Canada}

\author{G. Hagen}
\thanks{This manuscript has been authored by UT-Battelle, LLC under
  Contract No. DE-AC05-00OR22725 with the U.S. Department of
  Energy. The United States Government retains and the publisher, by
  accepting the article for publication, acknowledges that the United
  States Government retains a non-exclusive, paid-up, irrevocable,
  world-wide license to publish or reproduce the published form of
  this manuscript, or allow others to do so, for United States
  Government purposes. The Department of Energy will provide public
  access to these results of federally sponsored research in
  accordance with the DOE Public Access
  Plan. (http://energy.gov/downloads/doe-public-access-plan).}
\affiliation{Physics Division, Oak Ridge National Laboratory, Oak
  Ridge, Tennessee 37831, USA}
\affiliation{Department of Physics and Astronomy, University of
  Tennessee, Knoxville, Tennessee 37996, USA}

\author{T. Papenbrock}
\affiliation{Department of Physics and Astronomy, University of
  Tennessee, Knoxville, Tennessee 37996, USA}
\affiliation{Physics Division, Oak Ridge National Laboratory, Oak
  Ridge, Tennessee 37831, USA}

\date{\today}

\begin{abstract}
We compute the electric dipole polarizability of $^{48}$Ca with an
increased precision by including more correlations than in  previous
studies.  Employing the coupled-cluster method we go beyond singles
and doubles excitations and include leading-order
three-particle-three-hole ($3p$-$3h$) excitations for the ground
state, excited states, and the similarity transformed operator.  We
study electromagnetic sum rules, such as the bremsstrahlung sum rule
$m_0$ and the polarizability sum rule $\alpha_D$ using interactions
from chiral effective field theory.  To gauge the quality of our
coupled-cluster approximations we perform several benchmarks with the
effective interaction hyperspherical harmonics approach in $^4$He and
with self consistent Green's function in $^{16}$O.  We compute the
dipole polarizability of $^{48}$Ca employing the chiral interaction
N$^2$LO$_{\rm sat}$ [Ekstr\"om {\it et al.}, Phys. Rev. C {\bf 91},
  051301 (2015)] and the 1.8/2.0 (EM)~[Hebeler {\it et al.},
  Phys. Rev. C {\bf 83}, 031301 (2011)].  We find that the effect of
$3p$-$3h$ excitations in the ground state is small for 1.8/2.0 (EM)
but non-negligible for N$^2$LO$_{\rm sat}$. The addition of these new
correlations allows us to improve the precision of our $^{48}$Ca
calculations and reconcile the recently reported discrepancy between
coupled-cluster results based on these interactions and the
experimentally determined $\alpha_D$ from proton inelastic scattering
in $^{48}$Ca [Birkhan {\it et al.}, Phys. Rev. Lett. {\bf 118}, 252501
  (2017)].  For the computation of electromagnetic and polarizability
sum rules, the inclusion of leading-order $3p$-$3h$ excitations in the
ground state is important, while less so for the excited states and
the similarity-transformed dipole operator.
\end{abstract}


\maketitle

\section{Introduction}
\label{sec:intro}
%

The electric dipole polarizability $\alpha_D$ has been measured in
several nuclei~\cite{tamii2011,rossi2013,hashimoto2015,birkhan2017},
and is of key interest to theorists. On the one hand, mean-field
calculations suggest that $\alpha_D$ is strongly correlated to the
neutron skin, i.e. the difference between the root-mean-square (RMS)
radii of the neutron and proton
distributions~\cite{furnstahl2002,reinhard2010,roca-maza2011,piekarewicz2012,roca-maza2013,reinhard2013},
and thereby connects nuclei and neutron stars~\cite{tsang2012}. On the
other hand, computations of $^{48}$Ca by~\textcite{hagen2015} based on
Hamiltonians from chiral effective field
theory~\cite{epelbaum2009,machleidt2011} exhibit no correlations
between the neutron skin and $\alpha_D$, but do correlate the latter
with both the RMS radius of the neutron distribution and the symmetry
energy (and its slope) in nuclear matter, again providing a connection
between neutron-star physics and finite
nuclei~\cite{brown2000,horowitz2001}. The prediction~\cite{hagen2015}
and measurement~\cite{birkhan2017} of the dipole polarizability in
$^{48}$Ca agree within uncertainties, but theory somewhat
overestimates the experimental data for some interactions.  This
motivates us to revisit this nucleus in this work, with an emphasis on
computing $\alpha_D$ more precisely by going to the next level of
approximation and include more many-body correlations.

We note that the study of the role of many-body correlations, while
performed here with the coupled-cluster
method~\cite{kuemmel1978,bishop1991,mihaila2000b,dean2004,bartlett2007,binder2013b,hagen2014},
is relevant also for other methods such as self consistent Green's
functions~\cite{dickhoff2004}, in-medium similarity renormalization
group~\cite{tsukiyama2011,hergert2016}, and Gorkov-Green's function
approaches~\cite{soma2013}. All these methods seek economical ways to
include the necessary particle-hole correlations required to achieve a
precise calculation of energies and observables of ground- and excited
states~\cite{parzuchowski2017,parzuchowski2017b,Barbieri:2017hvp}. Thus,
we expect that our results will also be useful for those applications.

In the last decade, major advancements have been made in first
principles approaches to nuclear
structure~\cite{navratil2009,epelbaum2012,Leidemann:2012hr,barrett2013,hagen2014,carlson2015,hergert2016}. Realistic
descriptions of nuclei as heavy as $^{78}$Ni and $^{100}$Sn have
recently been achieved based on state-of-the-art nucleon-nucleon
($NN$) and three-nucleon forces (3NFs) from chiral effective field
theory~\cite{hagen2016b,simonis2017,morris2018}. This advancement is
based on the combination of first principles methods that scale
polynomial with system
size~\cite{mihaila2000b,dean2004,dickhoff2004,hagen2008,hagen2010b,tsukiyama2011,roth2012,hergert2013b,soma2013b,binder2013,lahde2014},
an ever-increasing computational power following Moore's law, insights
from the renormalization group and effective field
theory~\cite{bogner2003,bogner2007}, and
progress with nuclear forces~\cite{ekstrom2015,binder2015,carlsson2016,lynn2016}.

Electromagnetic reactions are a crucial tool to investigate nuclear
dynamics~\cite{Bacca2002,Gazit2006,Bacca2007,Efros2007}, see
Ref.~\cite{BaccaPastore2014} for a recent review.  Due to the
perturbative nature of the process, one can clearly separate the role
of the known electromagnetic probe from the less well known nuclear
dynamics. Through a comparison of experimental data with theory one is
then able to assess the precision and accuracy of the employed nuclear
interactions and associated current operators. Photo reactions on
heavier nuclei can be computed by the combination of the Lorentz
integral transform~\cite{efros1994} and the coupled-cluster
method~\cite{bacca2013,bacca2014,hagen2016}, and by alternative
means~\cite{Calci:2016dfb,Lovato:2016gkq,Barbieri:2017hvp,Stumpf:2017hiq,Rocco:2018vbf}.

A key ingredient to study electromagnetic reactions, such as
photo-dissociation or electron scattering, is the nuclear response
function, defined as
\begin{equation} 
R(\omega,q)\equiv\sum_{\mu}  \left|\langle \Psi_{\mu}| \hat{\Theta}(q)|\Psi_0\rangle \right|^2\delta\left(E_{\mu}-E_0-\omega \right).
\label{resp}
\end{equation}
Here $\omega$ is the transferred energy, while $\hat{\Theta}(q)$ is
the electromagnetic operator. It depends on the momentum-transfer $q$
of the considered probe. The nuclear response function is a dynamical
observables and requires knowledge of the ground state $|\Psi_0\rangle$
and all excited states $|\Psi_\mu\rangle$ with corresponding energies
$E_0$ and $E_\mu$, respectively. As most of the excited states are in
the continuum, the sum in Eq.~(\ref{resp}) really becomes an integral,
and this makes the direct computation of the response function a
formidable task.

Instead, it is often easier to compute sum rules of such response
functions, i.e. moments of the response intended as a distribution
function and defined as
\begin{equation} 
m_n(q)\equiv \int_0^{\infty} d\omega~ {\omega}^n R(\omega,q).
\label{sr}
\end{equation}
Here, $n$ is typically an integer.  Employing the closure relation, we
rewrite Eq.~(\ref{sr}) as a ground-state expectation value
\begin{equation} 
m_n= \langle \Psi_0 | \hat{\Theta}^{\dagger} (\hat{H} -E_0)^n \hat{\Theta}|\Psi_0 \rangle\,.
\label{sr_gs}
\end{equation}
In practice, one often inserts completeness relations on the right of
$\hat{\Theta}^{\dagger}$ and on the left of $\hat{\Theta}$, truncates
the employed Hilbert spaces, and then increases the number of states
until convergence is obtained.

While knowing the response function is equivalent to knowing all of
its (existing) moments, already from a few moments one can gain useful
insights into the dynamics of the nucleus.  In this paper we will take
this approach and focus on a few well known sum rules, namely the
bremsstrahlung sum rule $m_0$ and the polarizability sum rule
$\alpha_D\propto m_{-1}$ [see Eq.~(\ref{polresp}) below]. We calculate
these observables within coupled-cluster theory using various
approximation levels, and infer information on the nuclear dynamics
from a comparison to other available theoretical computations and/or
experimental data.

In the long wave-length approximation, i.e., in the limit
of $q \rightarrow 0$, the electric dipole operator reads
\begin{equation}
\label{dip}
\hat{ \Theta}=\sum_k^A \left({\bf r}_k - {\bf R}_{\rm cm}  \right) \left( \frac{1+\tau^3_k}{2} \right)\,.
\end{equation}
Here ${\bf r}_k$ and ${\bf R}_{\rm cm}$ are the coordinates of the
$k$-th nucleon and the center of mass of the nucleus, respectively,
and  $\tau^3_k$ is the third component of the isospin operator.
As an example, the photo-disintegration cross section of a
nucleus below the pion production threshold becomes
\begin{equation} 
\sigma_{\gamma}(\omega)=4\pi^2 \alpha \omega R(\omega)\,.
\label{cs}
\end{equation}
Here $R(\omega)$ is the response function~(\ref{resp}) of the
translationally invariant dipole operator~(\ref{dip}) for
$\omega=|{\bf q}|$.  Employing an $NN$ interaction from chiral
effective field theory~\cite{entem2003} this reaction cross section
was calculated in Refs.~\cite{bacca2013,bacca2014} using
coupled-cluster theory with singles and doubles excitations
(CCSD). While those calculations agreed with experimental data for
$^4$He, $^{16-22}$O and $^{40}$Ca, they lacked a better understanding
of uncertainties related to the underlying Hamiltonian and the applied
many-body approach.

The computation of a cross section or related sum rule exhibits
various theoretical uncertainties. Often, the largest uncertainty
results from truncations of the employed chiral effective field theory
at some given order in the power counting, whereas statistical
uncertainties due to the optimization of the interaction are
significantly
smaller~\cite{ekstrom2015b,perez2015,carlsson2016}. While a robust
quantification of systematic uncertainties related to the employed
chiral interactions is still lacking, a rough estimate can be made by
employing a large set of different chiral
interactions~\cite{hebeler2011,ekstrom2015,hagen2016,calci2016,hagen2016b}.
Uncertainties from finite harmonic-oscillator basis can be estimated
by varying the number of oscillator shells ($N_{\rm max}$) and the
oscillator frequency ($\hbar\Omega$). We found that this
uncertainty is of the order of $1\%$ for calculations in nuclei
with mass number up to $A=48$~\cite{hagen2016}. Another source of
uncertainty in the coupled-cluster method comes from
truncating the cluster operator at some low-order particle-hole excitation
rank. Previous works employed the CCSD approximation, and
neglected higher order excitations, such as $3p$-$3h$ excitations. It
is the purpose of this paper to investigate the role of leading-order
$3p$-$3h$ excitations in the ground state, excited states, and the
similarity transformed operator in the calculation of the
bremsstrahlung sum rule $m_0$~\cite{BSR1,BSR2,BSR3,BSR4} and on the
electric dipole polarizability sum rule $\alpha_D$~\cite{PSR}.

The electric dipole polarizability is related to the inverse energy
weighted sum rule $m_{-1}$ as
\begin{equation}\label{polresp}
\alpha_D = 2\alpha \int_{\omega_{ex}}^{\infty}  d\omega~\frac{R(\omega)}{\omega}\,=2\alpha~ m_{-1}.
\end{equation}
Due to the inverse energy weight, this sum rule is more sensitive to
the low-energy part of the excitation spectrum.

This paper is organized as follows. In Section~\ref{sec:th} we
introduce the formalism used to calculate the response functions, and
the sum rules $m_0$ and $\alpha_D$. We also present the nomenclature
used for the various approximation schemes that we implemented in this
work. In Section~\ref{res} we present benchmarking results for $m_0$
and $\alpha_D$ in $^4$He and $^{16}$O to validate our approach. In
Section~\ref{48}, we revisit our $^{48}$Ca calculations and compare
them with the recent experimental data of Ref.~\cite{birkhan2017}.
Finally, we draw our conclusions in Section~\ref{sec:conclusions}.

\section{Theoretical methods}
\label{sec:th}
Coupled-cluster
theory~\cite{coester1958,coester1960,kuemmel1978,mihaila2000b,dean2004,wloch2005,hagen2010b,binder2013b,hagen2014}
is based on the similarity transformed Hamiltonian,
\begin{equation} 
\overline{H}_N= e^{-T} H_N e^T, \;\; T = T_1 + T_2 + T_3 \ldots.
\end{equation} 
Here $H_N$ is normal-ordered with respect to a single-reference state
$\vert \Phi_0\rangle$ (usually the Hartree-Fock state), and $T$ is an
expansion in particle-hole excitations with respect to this
reference. The similarity transformation decouples all particle-hole
excitations from the ground state, and the reference state $\vert
\Phi_0\rangle $ becomes the exact ground state of $\overline{H}_N$. In
practice, the operator $T$ is truncated at some low rank particle-hole
excitation level. The similarity transformed Hamiltonian can be
evaluated using the Baker-Campbell-Hausdorff expansion, and terminates
exactly at quadruply nested commutators for a normal-ordered
Hamiltonian containing at most up to two-body terms. The drawback from
having an exactly terminating commutator expansion, is that the
similarity transformed Hamiltonian is non-Hermitian and thus requires
the computation of both the left and right eigenstates in order to
evaluate expectation values and transitions. The left ground state is
parameterized as 
\begin{equation}
  \langle \Psi_0 \vert =\langle \Phi_0 \vert (1+\Lambda), \;\; \Lambda = \Lambda_1 + \Lambda_2 + \Lambda_3 \ldots,
\end{equation}
where $\Lambda $ is a sum of particle-hole de-excitation
operators. The ground-state energy $E_0$ is given by the
energy-functional
\begin{equation} 
E_0 = E_{\rm HF} + \langle \Phi_0 \vert (1+\Lambda) \overline{H}_N \vert
\Phi_0 \rangle. 
\end{equation}
Here $E_{\rm HF} $ is the Hartree-Fock reference energy. The
left and right ground states are normalized according to $\langle
\Phi_0 \vert (1+\Lambda) \vert \Phi_0 \rangle = 1$.

For excited states we employ the equation-of-motion (EOM)
coupled-cluster method \cite{stanton1993} and calculate the right and
left excited state of $\overline{H}_N$, i.e.,
\begin{eqnarray} 
\label{eq:eomcc}
\nonumber
\overline{H}_N R_\mu \vert\Phi_0 \rangle = E_\mu R_\mu \vert \Phi_0
\rangle\, ,  \\
\langle \Phi_0 \vert L_\mu \overline{H}_N = E_\mu \langle \Phi_0\vert
L_\mu \, .
\end{eqnarray}
Here $R_\mu $ and $L_\mu $ are linear expansions in
particle-hole excitations with
\begin{eqnarray}
  \nonumber
  R_{\mu}&=& r_0+ \sum_{i,a}r^a_i \hat{a}^\dag_a \hat{a}_i +\frac{1}{(2!)^2}\sum_{i,j,a,b}r^{ab}_{ij}\hat{a}_a^\dag\hat{a}^\dag_b\hat{a}_j \hat{a}_i  \\
  &+&\frac{1}{(3!)^2}\sum_{i,j,k,a,b,c}r^{abc}_{ijk}\hat{a}_a^\dag\hat{a}^\dag_b \hat{a}^{\dag}_c \hat{a}_k \hat{a}_j \hat{a}_i \\
  \nonumber
  & +&\frac{1}{(4!)^2}\sum_{i,j,k,l,a,b,c,d}r^{abcd}_{ijkl}\hat{a}_a^\dag\hat{a}^\dag_b \hat{a}^{\dag}_c \hat{a}^{\dag}_d \hat{a}_l \hat{a}_k \hat{a}_j \hat{a}_i  \\
& + & \dots \,.
  \end{eqnarray}
The expression for $L_{\mu}$ is equivalent. Here, $\hat{a}^\dag$ and
$\hat{a}$ are creation and annihilation operators, respectively.
Indices $a,b,c,d$ run over unoccupied orbitals, while $i,j,k,l$ run
over occupied orbitals.  The left and right excited states are
normalized according to
\begin{equation}
\langle \Phi_0 \vert L_\mu R_{\mu'} \vert \Phi_0 \rangle =
\delta_{\mu,\mu'} \,.
\end{equation}

The electromagnetic transition strength from the ground to an excited
state is evaluated in coupled-cluster theory as 
\begin{eqnarray} 
 \vert \langle \Psi_{\mu} \vert \hat{\Theta}\vert \Psi_0 \rangle \vert^2 &  =  & \langle \Psi_0 \vert \hat{\Theta}^\dagger\vert \Psi_{\mu} \rangle \langle \Psi_{\mu} \vert \hat{\Theta}\vert \Psi_0 \rangle =\\
&=&
\nonumber
\langle \Phi_0 \vert (1 + \Lambda ) \overline{\Theta}_N^\dagger R_\mu \vert \Phi_0 \rangle 
\langle \Phi_0 \vert L_\mu \overline{\Theta}_N \vert \Phi_0 \rangle \,.
\end{eqnarray}
Here $\overline{\Theta}_N \equiv e^{-T}\Theta_Ne^T$ is the similarity
transformed normal-ordered operator, which in this work is taken to be
the electric dipole of Eq.~(\ref{dip}).  Because it is a one-body
operator, the Baker-Campbell-Hausdorff expansion $\overline{\Theta}_N$
terminates at doubly nested commutators
\begin{equation}
  \overline{\Theta}_N= \Theta_N + \left[ \Theta_N,T\right] + {1\over 2}\left[ \left[ \Theta_N,T\right],T\right]\,.
  \label{bhc}
\end{equation}

Having defined the ground and excited states, we can rewrite the
response function in the coupled-cluster formalism starting from
Eq.~(\ref{resp}) as
\begin{eqnarray}
\nonumber
R(\omega) & = & \sum_\mu \langle \Phi_0 \vert
(1+\Lambda) \overline{\Theta}_N^\dag R_\mu \vert \Phi_0 \rangle \langle 
\Phi_0 \vert L_\mu \overline{\Theta}_N \vert \Phi_0 \rangle \\
& \times & \delta( E_\mu - E_0 -\omega).
\label{noresp} 
\end{eqnarray}
By integrating over the energy $\omega $ we obtain the bremsstrahlung
sum rule $m_0$
\begin{eqnarray}
\label{eq:noresp} 
m_0 & = & \sum_\mu \langle \Phi_0 \vert
(1+\Lambda) \overline{\Theta}_N^\dag R_\mu \vert \Phi_0 \rangle \langle 
\Phi_0 \vert L_\mu \overline{\Theta}_N \vert \Phi_0 \rangle \ .
\end{eqnarray}
Using the closure relation we obtain the equivalent expression
\begin{equation}
 \label{oddo_compl}
 m_0=  \langle \Phi_0 \vert
(1+\Lambda) \overline{\Theta}_N^\dag \cdot \overline{\Theta}_N \vert \Phi_0 \rangle\,.
\end{equation}
In practice one solves for Eq.~(\ref{oddo_compl}) by inserting a complete
set on the Fock space defined by
\begin{equation} 
\label{projection}
1\!\!  = \!\!   \vert \Phi_0 \rangle\langle \Phi_0\vert +\sum_S\! \vert S\rangle\langle S\vert +
\sum_D\! \vert D\rangle \langle D\vert + \sum_T\! \vert T\rangle \langle T\vert + \sum_Q\! \vert
Q \rangle \langle Q\vert + \ldots . \\
\end{equation}
Here $S, D, T, Q$ label single, double, triple, and quadruple (and so
on) excited reference states, corresponding to
\begin{eqnarray}
  \label{eq:sdtq}
  \sum_S\vert S \rangle \langle S\vert & =&
  \sum_{ia}\vert \Phi_i^a \rangle \langle \Phi_i^a \vert \,,\\
  \nonumber
  \sum_D\vert D \rangle \langle D\vert& =&  \sum_{ijab}  \vert \Phi_{ij}^{ab} \rangle  \langle \Phi_{ij}^{ab}\vert  \,, \\
  \nonumber
  \sum_T\vert T \rangle \langle T\vert& =&
  \sum_{ijkabc}\vert \Phi_{ijk}^{abc} \rangle\langle \Phi_{ijk}^{abc}\vert \,, \\
  \nonumber
  \sum_Q\vert Q \rangle \langle Q\vert& =&\sum_{ijklabcd}\vert \Phi_{ijkl}^{abcd} \rangle\langle \Phi_{ijkl}^{abcd}\vert \,, \\
  \nonumber
  \dots &=& \dots \,.
\end{eqnarray}
Here, $\vert\Phi_{i_1\cdots i_n}^{a_1\cdots a_n}\rangle =
\hat{a}^\dagger_{a_1}\cdots\hat{a}^\dagger_{a_n}\hat{a}_{i_n}\cdots\hat{a}_{i_1}\vert\Phi_0\rangle$
is a $np$-$nh$ state.  Inserting the completeness~(\ref{projection})
into Eq.~(\ref{oddo_compl}) one obtains
\begin{eqnarray}
\label{eq:noresp2} 
m_0 & = & \langle \Phi_0 \vert
(1+\Lambda) \overline{\Theta}_N^\dag \vert  \Phi_0 \rangle \langle \Phi_0 \vert \overline{\Theta}_N\vert \Phi_0
\rangle\\
\nonumber
& = & \sum_S \langle \Phi_0 \vert (1+\Lambda) \overline{\Theta}_N^\dag \vert S\rangle\langle S \vert
\overline{\Theta}_N \vert \Phi_0 \rangle \\
\nonumber
& + &  
\sum_D \langle \Phi_0 \vert (1+\Lambda) \overline{\Theta}_N^\dag \vert D\rangle\langle D \vert
\overline{\Theta}_N \vert \Phi_0 \rangle \\
\nonumber
& + &  
\sum_T \langle \Phi_0 \vert (1+\Lambda) \overline{\Theta}_N^\dag \vert T\rangle\langle T \vert
\overline{\Theta}_N \vert \Phi_0 \rangle  \\
\nonumber
& + &  
\sum_Q \langle \Phi_0 \vert (1+\Lambda) \overline{\Theta}_N^\dag \vert Q\rangle\langle Q \vert
\overline{\Theta}_N \vert \Phi_0 \rangle + \ldots \,.
\end{eqnarray}
Here the first term is identically zero for non-scalar operators
$\hat{\Theta}$, such as the electric dipole operator considered in
this work. Calculating $m_0$ from Eq.~(\ref{eq:noresp2}) is
significantly simpler than starting from Eq.~(\ref{eq:noresp}) since
no knowledge of the excited states of $\overline{H}_N$ is required. As
a proof of principle we verified that solving Eq.~(\ref{eq:noresp2})
is equivalent to calculating the response function and integrating in
$\omega$ using Eq.~(\ref{eq:noresp}). 

The inverse-energy-weighted-polarizability sum rule $\alpha_D$ of
Eq.~(\ref{polresp}) can be calculated by utilizing the Lanczos
continued fraction method~\cite{miorelli2016} as
\begin{eqnarray}
  \label{eq:polresp}
  \nonumber
  \alpha_D & =& 2\alpha~ \langle \Phi_0\vert (1+\Lambda) \overline{\Theta}^\dag_N\frac{1}{\overline{H}_N - E_0}\overline{\Theta}_N\vert \Phi_0 \rangle \\
  & =  & 2\alpha ~\mathcal{I}_L(\sigma=0,\Gamma=0)\nonumber\\
  & =  & m_0\: x_{00}\,.
\end{eqnarray}
Here $\mathcal{I}_L(\sigma, \Gamma) $ is the Lorentz integral
transform, and $x_{00}$ is a continued fraction of the Lanczos
coefficients. To calculate the Lorentz integral
transform~\cite{efros1994,bacca2013}, one needs to solve an EOM with a
source term and the non-symmetric Lanczos algorithm is implemented by
constructing the right and left normalized pivots as (see
Ref.~\cite{miorelli2016} for details)
\begin{eqnarray}
\vert P_r\rangle & = & \overline{\Theta}_N\vert \Phi_0 \rangle, \\
\langle P_l\vert & = &  m_0^{-1} \: \langle \Phi_0\vert (1+\Lambda)
\overline{\Theta}^\dag_N, 
\end{eqnarray}
respectively.

So far we have not introduced any approximations in the
coupled-cluster formulations for ground and excited states.  The most
commonly used approximation for the ground state is CCSD (i.e. $T =
T_1 + T_2$, and $\Lambda = \Lambda_1 + \Lambda_2$), which typically
amounts for about $90\%$ of the full correlation energy in systems
with well defined single-reference character~\cite{bartlett2007}. In
the following we will denote the CCSD approximation in short with D.
We will also go beyond the CCSD level by including leading-order
$3p$-$3h$ excitations using the CCSDT-1
approach~\cite{lee1984}. CCSDT-1 is a good approximation to the full
CCSDT approach and accounts for about $99\%$ of the correlation
energy. In brief, CCSDT-1 is an iterative approach that includes the
leading-order contribution $\left(H_NT_2\right)_C$ (here the index $C$
denotes that only connected terms
contribute~\cite{shavittbartlett2009}) to the $T_3$ amplitudes with an
energy denominator given by the Hartree-Fock single-particle energies,
while all $T_3$ contributions to the $T_1$ and $T_2$ amplitudes are
fully included. We will also solve for the corresponding left ground
state in the CCSDT-1 following Ref.~\cite{watts1995}.  To simplify the
notation, in the following we will label the CCSDT-1 approximation
with T-1.  The corresponding approximations we will employ for excited
states given in Eq.~(\ref{eq:eomcc}) includes up to $2p$-$2h$ in the
EOM-CCSD approach, and leading-order $3p$-$3h$ excitations in the
EOM-CCSDT-1 approach~\cite{watts1995,jansen2016}. Because the
calculation of $m_0$ and $\alpha_D$ requires a particle-hole expansion
of the ground state ($T$ and $\Lambda$) and one for excited states
[$R_{\mu}$ and $ L_{\mu}$ for $\alpha_D$ or Eq.~(\ref{eq:noresp2}) for
  $m_0$], we need to label both of them appropriately.  In this work
we will investigate different approximation levels in both the ground
and excited states. In order to keep the notation concise we therefore
denote each scheme with a pair of labels (separated by a `$/$'
symbol), with the largest order of correlation included in the ground
state on the left, and the largest order of correlation included in
the excited states on the right as shown in Table~\ref{table:labels}.
In the previous work on dipole strengths and
polarizabilities~\cite{bacca2013,bacca2014,hagen2016,miorelli2016,birkhan2017}
both ground and excited states were approximated at the CCSD level, an
approximation we label by D/D in this work.

\begin{table}[h]
\caption{List of labels to denote the various coupled-cluster
  expansions used for the ground state (left of `$/$') and for the
  excited states (right of `$/$').  The symbol (S for singles, D for
  doubles, T for triples and T-1 for linearized triples) represents
  the highest order of correlation considered (with all lower orders
  always fully included).}
\label{table:labels}
\begin{center}
\begin{tabular}{lll}
\hline\hline
ground state~~ & EOM~~ & label~~\\
\hline
D & S   & D/S \\
D & D &  D/D \\
T-1 & S & T-1/S\\
T-1 & D & T-1/D\\
T-1 & T-1 & T-1/T-1\\
\hline
\end{tabular}
\end{center}
\end{table}

Examining the similarity transformed one-body operator
$\overline{\Theta}_N$ of Eq.~(\ref{bhc}) reveals that for $T=T_1 +
T_2$, the expansions of Eqs.~(\ref{eq:noresp2}) and (\ref{eq:polresp})
terminate at triply excited determinants ($\vert T\rangle $). If one
includes $T=T_1 + T_2 + T_3$ the expansions terminate at quadruply
excited determinants. Thus, the inclusion of $3p$-$3h$ excitations in
the ground and excited states requires the implementation of a number
of new coupled-cluster diagrams. As usual, we checked such diagrams by
comparing the $j$-coupling and $m$-coupling schemes (see, e.g.,
Ref.~\cite{Mirko_Thesis}).


\section{Validation and benchmarking}
\label{res} 

For a validation of our approach, we benchmark our results on $^{4}$He
and $^{16}$O. In all the results presented in this Section, $^4$He is
calculated with the the chiral $NN$ interaction at
next-to-next-to-next-to-leading order (N$^3$LO) from
Ref.~\cite{entem2003}, and $^{16}$O is calculated using the
N$^2$LO$_{\rm sat}$ interaction~\cite{ekstrom2015}, respectively.  The
choice of interaction for $^4$He is motivated by the fact that we want
to benchmark the various approximation schemes against virtually exact
results from the effective hyperspherical harmonics
approach~\cite{barnea2001}, which cannot easily employ the
N$^2$LO$_{\rm sat}$ due to the non-locality of the 3NF. For $^{16}$O,
the inclusion of 3NFs is necessary for a realistic description of the
charge radius and observables correlated with it, such as $m_0$ and
$\alpha_D$\cite{miorelli2016,hagen2016,birkhan2017}. This makes the
interaction N$^2$LO$_{\rm sat}$ a good choice. Using this interaction
also allows us to benchmark with self consistent Green's function
results~\cite{Barbieri:2017hvp}.

We first explore how the similarity-transformed transition operator
depends on the truncation level of included triples excitations. Then
we compute the $m_0$ sum rule and the dipole polarizability for $^4$He
and $^{16}$O. For $^{16}$O we also show a comparison of running sum
rules and discretized responses, which allow us to monitor how excited
states move as a function of energy for the various approximation
schemes.

\subsection{The similarity transformed transition operator}
In the T-1/S, T-1/D, and T-1/T-1 approaches, $T_3$ contributions are
included in the one- and two-body parts of the similarity transformed
Hamiltonian, while three-body parts from $T_3$ are only included
via $\left( F_NT_3 \right)_C$ (here $F_N$ is the normal ordered
one-body Fock matrix). By treating the similarity transformation of a
normal-ordered one-body operator $\hat{\Theta}_N$ consistently with
the similarity-transformed Hamiltonian, one has
\begin{eqnarray}
  \nonumber
  \overline{\Theta}_N&=&\left [ \Theta_N e^{T_1+T_2+T_3}\right]_C= \\
\label{oddo1}
  &=& \overline{\Theta}^D_N + \left[ \Theta_N \!\left(\!\frac{T_2^2}{2}+T_3+T_1T_3 \right) \right ]_C\\
  \label{oddo2}
                  &                                  \simeq &\overline{\Theta}^D_N + \left[ \Theta_N \left( \frac{T_2^2}{2}\right) \right ]_C\\
\label{oddo3}
	& \simeq &\overline{\Theta}^D_N\ ,
  \end{eqnarray}
where the $C$ index again denotes connected
diagrams~\cite{shavittbartlett2009} and $\overline{\Theta}^D_N$ is the
similarity-transformed operator in the D approximation.  Due to the
hierarchy among correlations, one can expect that the terms in
Eq.~(\ref{oddo1}) that contain $T_3$ are sub-leading with respect to
the $T^2_2$ term. These terms are also computationally much more
demanding since they involve calculating and storing $T_3$
configurations (see Ref.~\cite{Mirko_Thesis} for full
expressions). Thus, it is convenient to explore their relevance with
respect to using Eq.~(\ref{oddo2}), or even just using
Eq.~(\ref{oddo3}), where the operator is similarity transformed as in
the D approximation.

In this paper we compute observables by including $3p$-$3h$
excitations in the ground and excited states as well as in the
similarity transformed operator, and benchmark the various
approximations for $\overline{\Theta}_N$ in $^4$He and $^{16}$O, as
shown in Table~\ref{table:checkextradiag}. We see that for both $m_0$
and $\alpha_D$, the additional terms in Eqs.~(\ref{oddo1}) and
(\ref{oddo2}) have a negligible contribution with respect to
Eq.~(\ref{oddo3}), amounting to a sub-percent effect of about 0.2 and
0.7\%, respectively.  This finding is important in the light of
performing computations of heavy nuclei, where calculations with
Eq.~(\ref{oddo3}) are more tractable, while using Eq.~(\ref{oddo1})
would be substantially more computationally demanding. Consequently,
when calculating $^{48}$Ca in Section~\ref{48}, we will use Eq.~(\ref{oddo3}).

\begin{table}[h]
  \caption{Effect of contributions of $3p$-$3h$ correlations in a
    T-1/T-1 computation, when $\overline{\Theta}_N$ is truncated as in
    Eqs.~(\ref{oddo1}) or (\ref{oddo2}), with respect to the full
    expression of Eq.~(\ref{oddo3}). Both $m_0$ and $\alpha_D$ are
    computed with (a) $\hbar\Omega=26$ MeV and $N_{\rm max}=14$ and
    (b) $\hbar\Omega=22$ MeV, $N_{\rm max}=12$ and $E^F_{\rm
      3max}=14$.}
\label{table:checkextradiag}
\begin{center}
\begin{tabular}{c c c  c}
\hline\hline
 & ~~$^4$He$^{(a)}$~~ &~~ $^{16}$O$^{(b)}$~~ &  scheme\\
\hline
$m_0{\rm [fm^2]}$ & $\begin{matrix}
  0.951 \\[0.5pt]
  0.950 \\[0.5pt]
  0.949 
\end{matrix}$ &  $\begin{matrix}
  4.87\ \\[0.5pt]
  4.92\\[0.5pt]
  4.90
\end{matrix}$ & $\begin{matrix}
  {\rm ~Eq.~(\ref{oddo1})} \\[0.5pt]
    {\rm ~Eq.~(\ref{oddo2})}\\[0.5pt]
    {\rm ~Eq.~(\ref{oddo3})}
\end{matrix}$\\
\hline
$\alpha_D{\rm [fm^3]}$ &  $\begin{matrix}
  0.0816 \\[0.5pt]
  0.0808 \\[0.5pt]
  0.0811
\end{matrix}$ &
$\begin{matrix}
  0.523 \\[0.5pt]
  0.528 \\[0.5pt]
  0.527
\end{matrix}$&  $\begin{matrix}
  {\rm ~Eq.~(\ref{oddo1})} \\[0.5pt]
    {\rm ~Eq.~(\ref{oddo2})} \\[0.5pt]
    {\rm ~Eq.~(\ref{oddo3})}
\end{matrix}$\\
\hline
\end{tabular}
\end{center}
\end{table}

Note that when using Eq.~(\ref{oddo3}) in the calculation of $m_0$ in
the T-1/T-1 approximation, $3p$-$3h$ excitations enter only in the
ground state [see Eq.~(\ref{oddo_compl})], and therefore this
corresponds to the T-1/D approximation. On the contrary, triples would
enter both in the ground and excited states in a calculation of
$\alpha_D$, for which T-1/T-1 and T-1/D are different.

\subsection{$^4$He}

We now focus on $^4$He and explore the convergence in terms of the
model space size $N_{\rm max}$ for two approximation schemes that
includes $3p$-$3h$ excitations, namely T-1/T-1 and T-1/D, and compare
it to the D/D approximation.

\begin{figure}[th]
	\begin{center}
		\includegraphics[width=0.95\linewidth]{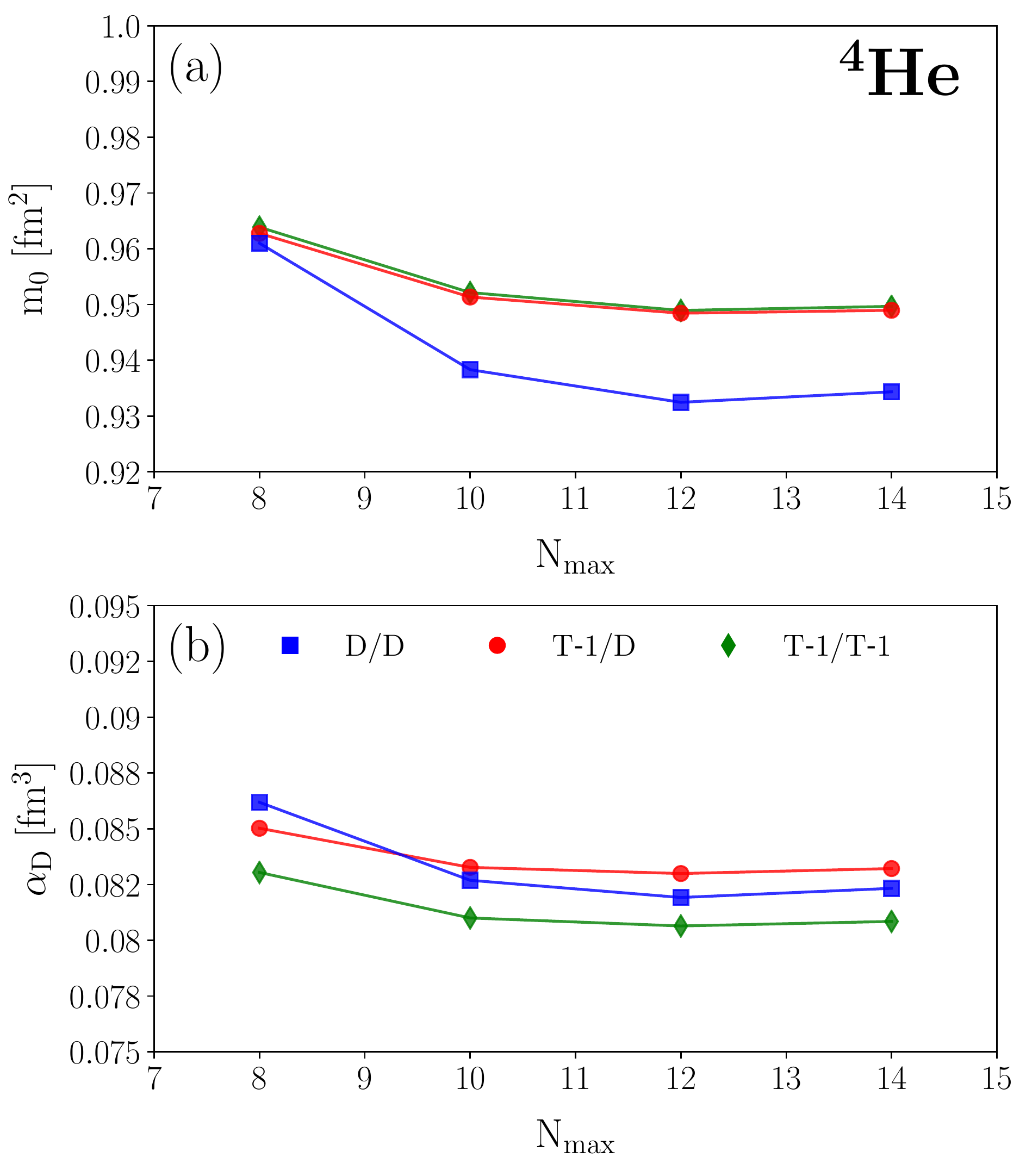}
    		\caption{(Color online) Convergence of $m_0$ (a) and
                  $\alpha_D$ (b) for $^4$He with respect to the
                  model-space size $N_{\rm max}$ for $\hbar\Omega=26$
                  MeV.  Two triples approximations schemes T-1/T-1
                  (green diamonds) and T-1/D (red circles) are
                  compared to the D/D case (blue squares).  The
                  similarity transformed operator is implemented with
                  Eq.~(\ref{oddo2}) and (\ref{oddo3}), in T-1/T-1 and
                  T-1/D, respectively.}
      \label{fig:4He_conv_CCSDT1_hw}
  \end{center}
\end{figure}

Fig.~\ref{fig:4He_conv_CCSDT1_hw} shows the convergence of $m_0$ and
$\alpha_D$ in $^4$He with respect to $N_{\rm max}$ for
$\hbar\Omega=26$~MeV.  Calculations in the T-1/T-1 and T-1/D scheme
were performed with Eq.~(\ref{oddo2}) and Eq.~(\ref{oddo3}),
respectively. The convergence with respect to $N_{\rm max}$ is of
similar quality both for $m_0$ (a) and $\alpha_D$ (b).

We see that for $m_0$, the results obtained within the T-1/D
approximation are close to those obtained in T-1/T-1
approximation. The slight difference stems from the fact that the
T-1/T-1 calculations are performed with the similarity transformed
operator given in Eq.~(\ref{oddo2}), while T-1/D results are obtained
with Eq.~(\ref{oddo3}). We implement these two different equations to
graphically show that our findings presented in
Table~\ref{table:checkextradiag} are consistent in various model
spaces and approximations.  For $\alpha_D$ we observe a slightly
larger difference between the T-1/T-1 and T-1/D approximations, due to
the fact that $3p$-$3h$ excitations enter in the calculations of
excited states as well. Overall, the effects of $3p$-$3h$ excitations
in $^4$He are small, amounting to about 1.5$\%$.

\begin{figure}[th]
	\begin{center}
		\includegraphics[width=0.95\linewidth]{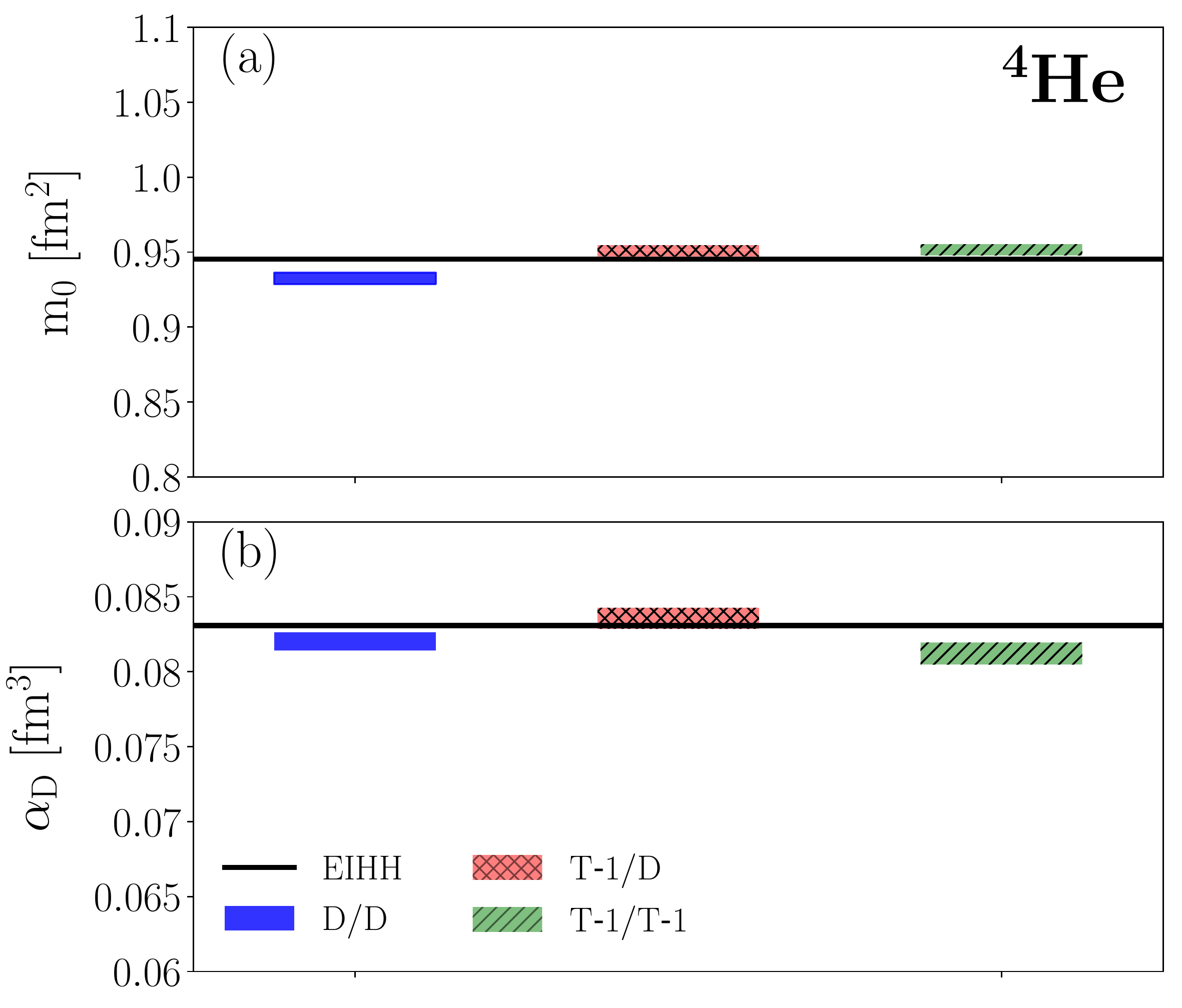}
    		\caption{(Color online) Comparison of $m_0$ (a) and
                  $\alpha_D$ (b) in the D/D (blue/left), the
                  T-1/D (red/central) and the T-1/T-1 (green/right)
                  approximations against hyperspherical harmonics
                  results (black line) in $^4$He. }
      \label{fig:m0_pol_bands}
  \end{center}
\end{figure}

It is now interesting to compare the various coupled-cluster results
with respect to the hyperspherical harmonics benchmark
values~\cite{barnea2001,bacca2013} for $^4$He.
Figure~\ref{fig:m0_pol_bands} shows $m_0$ (a) and $\alpha_D$ (b)
obtained for various approximations in coupled-cluster theory: D/D
(blue/left), the T-1/D (red/central) and the T-1/T-1
(green/right). The widths of the bands reflect the residual
$\hbar\Omega$ dependence for the largest model space $N_{\rm
  max}=14$. The black line is the virtually exact calculation from
hyperspherical harmonics expansions using the same interaction.  The
D/D calculations already get close to the hyperspherical harmonics
result, and the addition of $3p$-$3h$ correlations in both the T-1/D
and T-1/T-1 approaches further improves the agreement for $m_0$. For
$\alpha_D$ the T-1/D calculation agrees better with the hyperspherical
harmonics result than the T-1/T-1 approach.  The overall effect of
$3p$-$3h$ excitations is small. This benchmark with hyperspherical
harmonics suggests that the T-1/D scheme is to be preferred for
electromagnetic and polarizability sum rules, and that the inclusion
of $3p$-$3h$ correlations in the ground state plays a more significant
role than the corresponding one in the excited states.

\subsection{$^{16}$O}

\begin{figure}[h]
	\begin{center}
		\includegraphics[width=\linewidth]{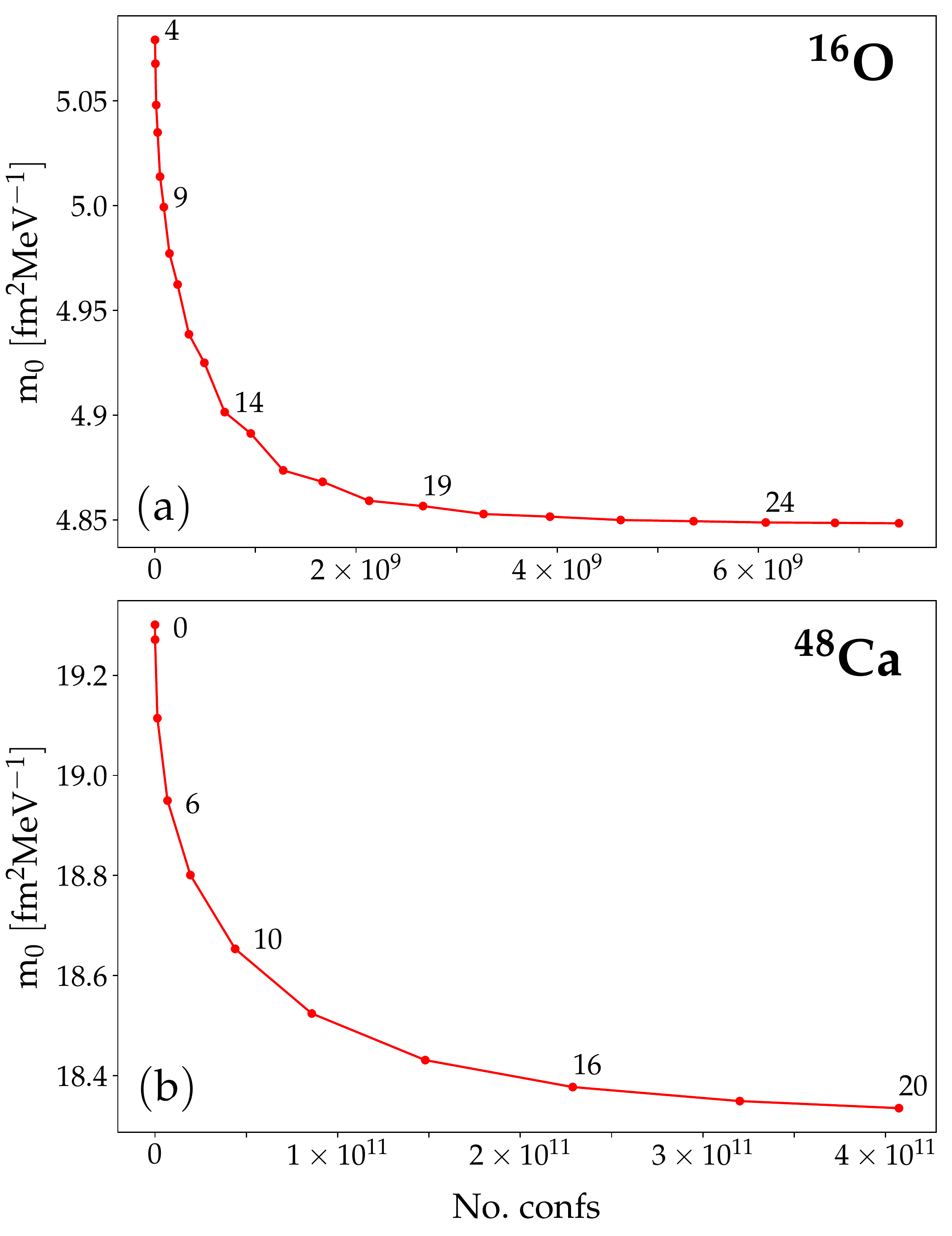}
    		\caption{(Color online) Convergence of $m_0$ for
                  $^{16}$O  (a) and $^{48}$Ca (b)
                  calculated in the T-1/D scheme as a function of the
                  number of configurations for the $E^F_{\rm
                    3max}$. The adopted model space is $N_{\rm
                    max}=12$. Each point corresponds to a jump of one
                  unit (two units) in $E^F_{\rm 3max}$ for $^{16}$)
                  ($^{40}$Ca) and selected values of $E^F_{\rm 3max}$
                  are highlighted along the curves.}
      \label{fig:E3max_curve}
  \end{center}
\end{figure}

Let us turn to $^{16}$O. First, we check the convergence of the $m_0$
sum rule with respect to the number of $3p$-$3h$ configurations
included in our calculations.  Going beyond the D/D approximation and
including $3p$-$3h$ excitations in the T-1/S, T-1/D, and T-1/T-1
approaches, the computational cost grows significantly both in terms
of number of computational cycles and memory associated with storage
of the amplitudes. The computational cost associated with the most
expensive term in the D/D approximation is given by $n_o^2 n_u^4$,
while in the T-1/S, T-1/D, and T-1/T-1 approaches it is $n_o^3
n_u^4$. Here $n_o$ is the number of occupied orbitals in the reference
state $|\Phi_0\rangle $ and $n_u$ is the number of unoccupied
orbitals. Clearly, the computational load grows rapidly with the mass
of the nucleus ($n_o=A$) and the model-space size ($n_u\propto N_{\rm
  max}^3$). In order to overcome this computational hurdle in the
T-1/S, T-1/D, and T-1/T-1 approaches, we introduce an energy cut
$E^F_{\rm 3max}$ on the allowed $3p$-$3h$ excitations. The truncated
space that we employ for the $3p$-$3h$ excitations are thus given by
$\vert N_a - N_F\vert + \vert N_b - N_F\vert + \vert N_c - N_F\vert
\leq E^F_{\rm 3max}$ and $\vert N_i - N_F\vert + \vert N_j - N_F\vert
+ \vert N_k - N_F\vert \leq E^F_{\rm 3max}$ with $N_p = 2n_p+l_p$
being the harmonic-oscillator shell and $N_F$ the harmonic-oscillator
shell at the Fermi surface. The top of Fig.~\ref{fig:E3max_curve}
shows the convergence with respect to $E^F_{\rm 3max}$ for $m_0$ in
$^{16}$O.  We observe that truncating $E^F_{\rm 3max}$ to 14 yield
results for $^{16}$O that are converged at the 1$\%$-level. Unless
stated otherwise, in the remainder of this work we will use $E^F_{\rm
  3max}=14$. Note that this truncation also works well for $^{48}$Ca,
as shown in the bottom part of Fig.~\ref{fig:E3max_curve}.

\begin{figure}[th]
	\begin{center}
		\includegraphics[width=0.95\linewidth]{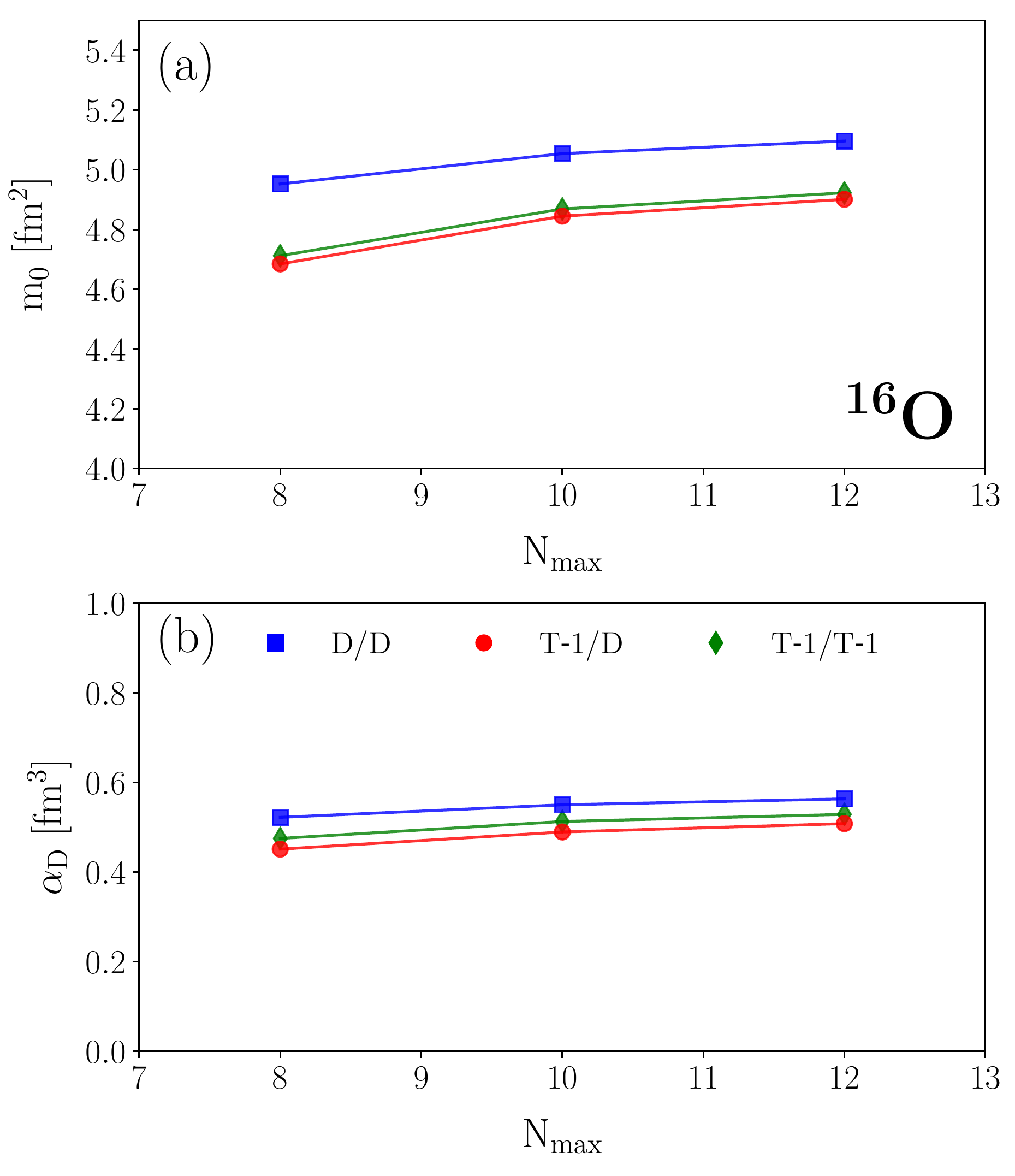}
    		\caption{(Color online) Convergence of $m_0$ (a) and
                  $\alpha_D$ (b) for $^{16}$O with respect to
                  the model-space size $N_{\rm max}$ for
                  $\hbar\Omega=22$ MeV and $E^F_{\rm 3max}=14$. Two
                  triples approximations schemes T-1/T-1 (green
                  diamonds) and T-1/D (red circles) are compared to
                  the D/D case (blue squares).  The similarity
                  transformed operator is implemented with
                  Eq.~(\ref{oddo2}) and (\ref{oddo3}), in T-1/T-1 and
                  T-1/D, respectively. }
      \label{fig:16O_conv_CCSDT1_hw}
  \end{center}
\end{figure}

In Fig.~\ref{fig:16O_conv_CCSDT1_hw} we show results for the $m_0$ sum
rule (top) and the dipole polarizability (bottom) of $^{16}$O at
$\hbar\Omega=22$~MeV and $E^F_{\rm 3max}=14$ as a function of $N_{\rm
  max}$. For $m_0$, the T-1/T-1 and T-1/D calculations almost
coincide, while some difference is observed in $\alpha_D$. The
residual $\hbar\Omega$ dependence amounts to about 1.5\% in the
largest model space.  While in $^{16}$O the effect of triples is
slightly larger than in $^{4}$He, the overall effect of $3p$-$3h$
excitations is small and amounts to 4$\%$ and 6$\%$ for $m_0$ and
$\alpha_D$, respectively.  Both for $m_0$ and $\alpha_D$ the inclusion
of $3p$-$3h$ excitations in the T-1/T-1 and T-1/D approaches reduce
their magnitude as compared to the results obtained in the D/D
approach.

\begin{figure}
	\begin{center}
\includegraphics[width=\linewidth]{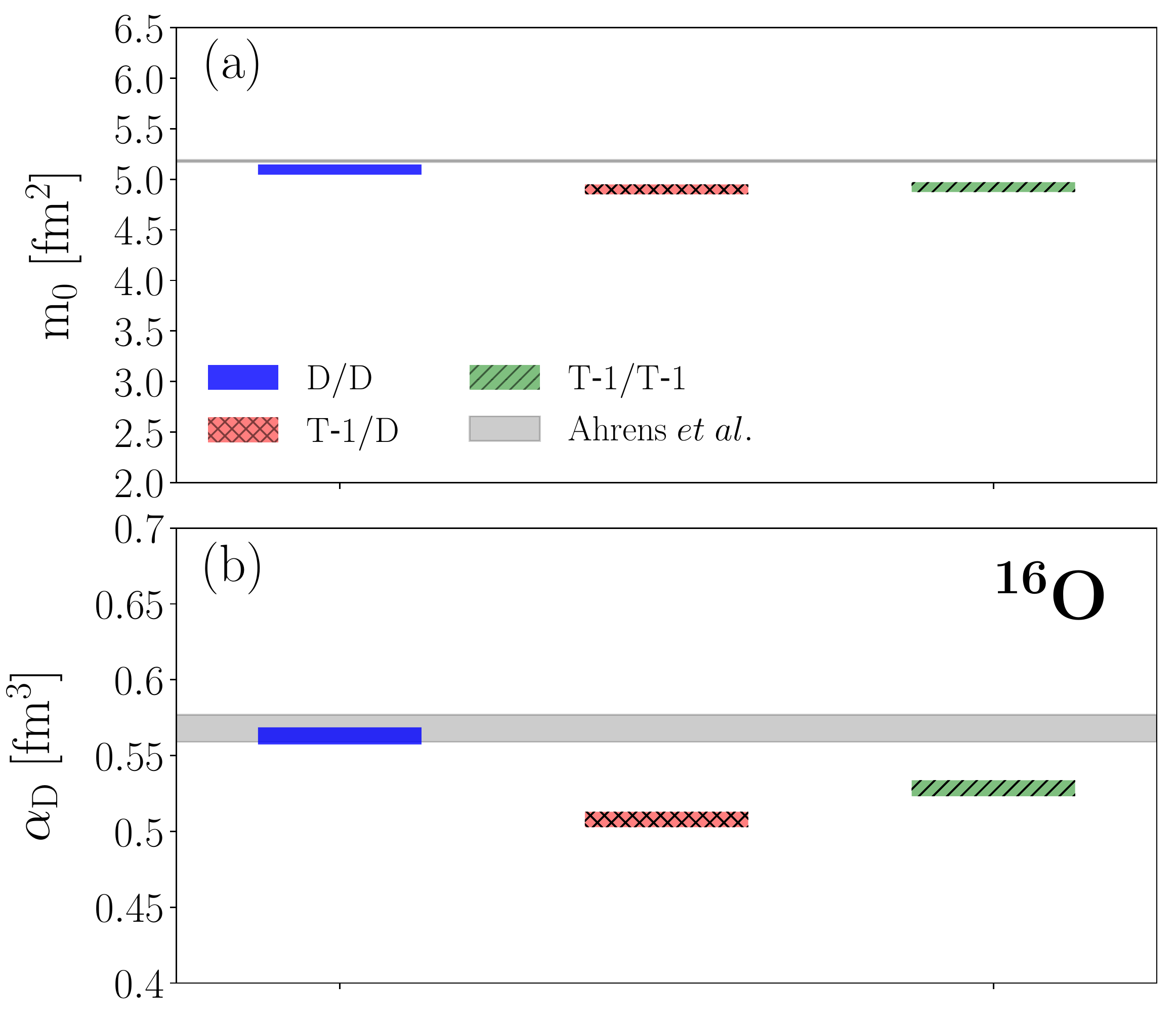}
\caption{(Color online) Comparison of $m_0$ (a) and $\alpha_D$
  (b) in the D/D (blue/left), the T-1/D (red/central) and the
  T-1/T-1 (green/right) approximations against the integrated
  experimental data by \textcite{Ahrens1975} in
  $^{16}$O.}
      \label{fig:16O_band}
  \end{center}
\end{figure}

In Fig.~\ref{fig:16O_band} we compare $m_0$ (a) and $\alpha_D$
(b) for $^{16}$O obtained in the D/D (blue/right) scheme with the
T-1/D (red/central) and the T-1/T-1 (green/right) approximations. The
D/D value is obtained at $N_{\rm max}=14$ and $\hbar\Omega=22$ MeV.
Due to the large number of $3p$-$3h$ configurations, for $^{16}$O we
are able to calculate only up to a maximum model space of $N_{\rm
  max}=12$ and $E^F_{\rm 3max}=14$ for T-1/T-1. For consistent results
we adopt the same truncation for $E^F_{\rm 3max}$ in T-1/D. The bands
in Fig.~\ref{fig:16O_band} are obtained by assigning a $2\%$
uncertainty, accounting for the combined uncertainty from the
$E^F_{\rm 3max}$ cut and the residual $\hbar \Omega$-dependence.

Correlations arising from $3p$-$3h$ excitations reduce the size of
these observables by a few percent, with effects being slightly larger
on $\alpha_D$ than for $m_0$. Similar to the $^4$He case, we find that
results for $m_0$ obtained in the T-1/D and T-1/T-1 approaches almost
coincide, while for $\alpha_D$ they slightly differ. This is expected,
because $m_0$ is calculated as a ground-state expectation value, while
$\alpha_D$ also requires the solution of the excited states from
Eq.~(\ref{eq:eomcc}). We use the difference between the T-1/D and
T-1/T-1 results as an estimate of neglected higher order
correlations. This amounts to a 4$\%$ effect for $\alpha_D$ and an
even smaller effect of 0.4$\%$ for $m_0$.

We now confront our results with data. Figure~\ref{fig:16O_band} compares
the theoretical results with the experimental value obtained by
integrating the data from \textcite{Ahrens1975} (grey bands). We see
that the addition of triples leads to a deviation of $m_0$ and
$\alpha_D$ with respect to the experimental data, which agreed better
in the D/D approximation using the N$^2$LO$_{\rm sat}$ interaction. We
note that N$^2$LO$_{\rm sat}$ was constrained to reproduce the charge
radius of $^{16}$O using coupled-cluster theory in the D
approximation. As shown in Refs.~\cite{hagen2015,miorelli2016}, the
charge radius is correlated with $\alpha_D$, and it would therefore be
interesting to quantify the effect of $3p$-$3h$ excitations in the T-1
approach on charge radii as well.  We also note that the extraction of
these sum rules from photo-absorption data may be prone to larger
systematic uncertainties than those quoted because it is not possible
to estimate the role of multipoles beyond the dipole.

As the experimental determination of the dipole polarizabilty results
from an integration of the dipole strength, it is interesting to study
the running of the $\alpha_D$ sum rules as a function of the maximum
integration limit. If one solves for the excited states in
Eq.~(\ref{eq:eomcc}) using the Lanczos technique, it is possible to
define the sum rule from the integral of the dipole response function
$R(\omega)$ as
\begin{equation} 
m_n(\varepsilon)= \int_0^{\varepsilon} d\omega~ {\omega}^n R(\omega)\,,
\label{running_sr}
\end{equation}
and study its running as a function of $\varepsilon$.  A discretized
response function can also be obtained by a calculation of the Lorentz
integral transform ${\mathcal
  I}_L(\sigma,\Gamma)$~\cite{efros1994,bacca2013} for a very small
width parameter $\Gamma$ as (see Ref.~\cite{miorelli2016} for details)
\begin{equation} 
m_n(\varepsilon)= \lim_{\Gamma \to 0} \int_0^{\varepsilon} d\sigma~ {\sigma}^n {\mathcal I}_L(\sigma,\Gamma)\,.
\label{running_sr_L}
\end{equation}
The discretized response consists of smeared $\delta$ peaks and does not
properly take the continuum into account. However, it will allow us to
see how excited states and their corresponding strengths change within
various approximation schemes, thus affecting the running sums.

\begin{figure}[t]
	\begin{center}
\includegraphics[width=1.0\linewidth]{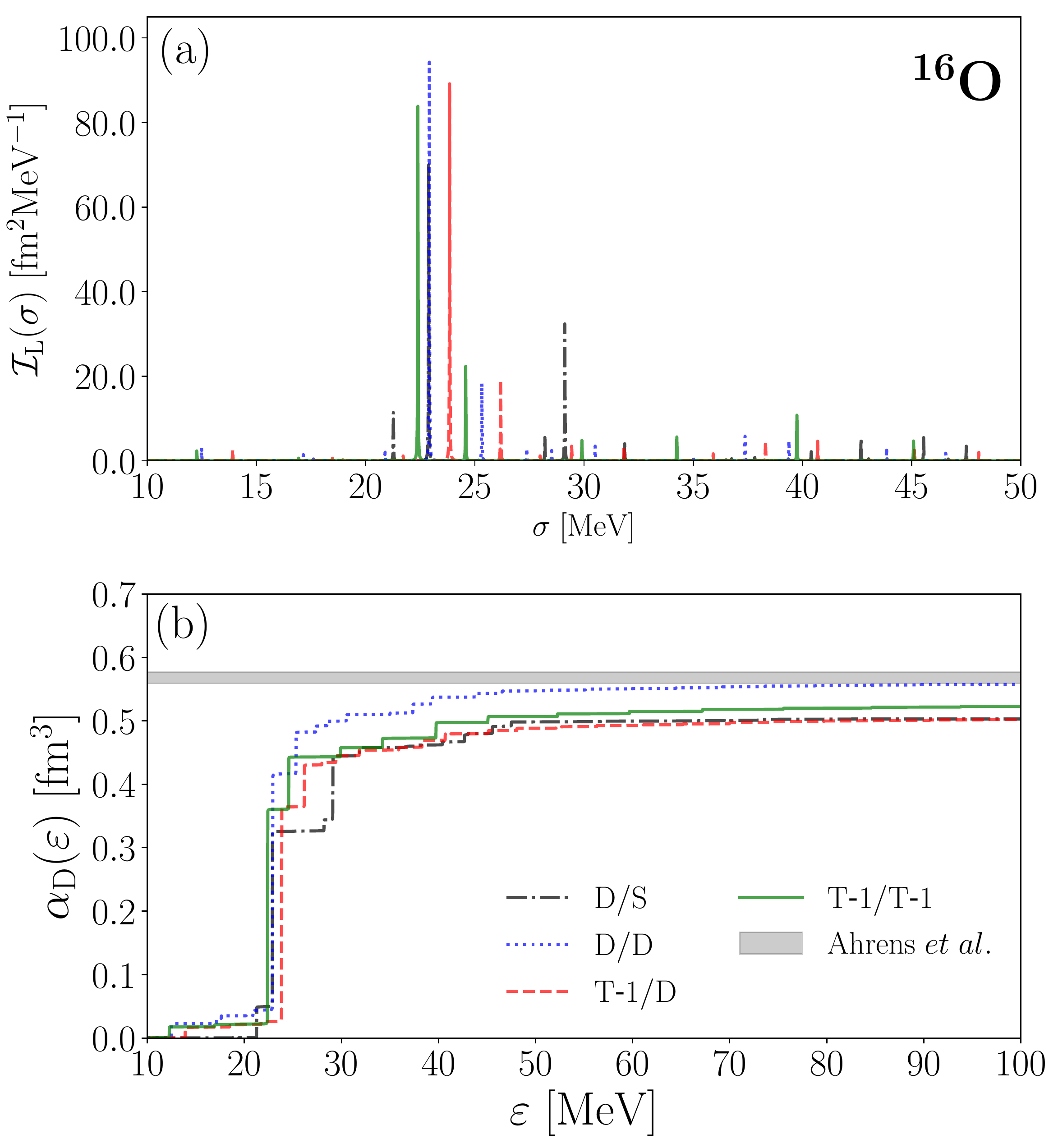}
\caption{(Color online) Discretized response with $\Gamma=0.01$ MeV
  (a) and running sum of $\alpha_D(\varepsilon)$ (b)
  for $^{16}$O with various coupled-cluster approximations compared
  with experimental results from \textcite{Ahrens1975}.}
      \label{fig:16O_running}
  \end{center}
\end{figure}

The top of Fig.~\ref{fig:16O_running} shows the Lorentz integral
transform calculated for $\Gamma=0.01$ MeV using various
coupled-cluster approximations for $^{16}$O.  The bottom plot shows
the running of the polarizability sum rule as a function of
$\varepsilon$. The various approximation schemes are shown in
comparison with experimental data (gray bands). Besides the D/D
calculation, we present three different schemes with increasing
correlation order, namely D/S, T-1/D and T-1/T-1.  We note that the
D/S scheme coincides with the T-1/D calculation, and is also very
close to the most expensive T-1/T-1 calculation, deviating from the
D/D approximation by a few percent.

\begin{table}[h]
\caption{Comparison of coupled-cluster results with self consistent
  Green's function data from Ref.~\cite{Barbieri:2017hvp} in $^{16}$O
  using the N$^2$LO$_{\rm sat}$ interaction.}
\label{comp_carlo}
\begin{center}
\begin{tabular}{lc}
\hline\hline
Method~~ & ~~$\alpha_D$ [fm$^3]$ \\
\hline
Ref.~\cite{Barbieri:2017hvp} & 0.50~~ \\
D/S &  0.503\\
T-1/D & 0.508 \\
T-1/T-1 & 0.528 \\
\hline
\end{tabular}
\end{center}
\end{table}

In Table~\ref{comp_carlo} we compare our results for $\alpha_D$ with
those of \textcite{Barbieri:2017hvp}. We see that the results obtained
with the D/S approach agree with calculations from the self consistent
Green's functions method that used a random phase approximation
approach for the excited states. We find that the inclusion of
$3p$-$3h$ correlations in the ground state and in the excited states
decreases the polarizability with respect to a D/D calculation, making
this more sophisticated calculation to coincidentally agree with
simpler schemes, such as the D/S approximation.

\section{Electric dipole polarizability of  $^{48}$Ca}
\label{48}

We now turn to our main goal and compute the dipole polarizability of
$^{48}$Ca. In the recent experiment using proton inelastic scattering
off a $^{48}$Ca target, the electric dipole strength was disentangled
from other multipole contributions resulting in
$\alpha_D=2.07(22)$~fm$^3$~\cite{birkhan2017}. Overall, theory agreed
with experiment, given the (still significant) systematic theoretical
uncertainties as one varies the employed interaction. However,
calculations based on interactions that reproduced the charge radius
of $^{48}$Ca yielded results for $\alpha_D$ in the D/D approximation
that were somewhat larger than the measured value. Those calculations
were performed in the D/D approximation, and this makes it interesting
to compute $\alpha_D$ with an increased precision by including
leading-order $3p$-$3h$ correlations in the T-1 approximation.

Taking advantage of the results and benchmarks from Section~\ref{res},
we revisit these calculations by using two established interactions,
namely N$^2$LO$_{\rm sat}$ and 1.8/2.0 (EM).  These two interactions
were also used in Refs.~\cite{hagen2015,birkhan2017}.  The 1.8/2.0
(EM) potential is constructed following Ref.~\cite{nogga2004}. It
starts from the chiral $NN$ interaction at N$^3$LO~\cite{entem2003}
and ``softens'' it with the similarity renormalization group
\cite{bogner2007} at a cutoff/resolution scale $\lambda_{\rm SRG} =
1.8~\rm{fm}^{-1}$. The 3NF is taken as the leading chiral 3NF using a
non-local regulator and cutoff $\lambda_{\rm 3NF} = 2.0~\rm{fm}^{-1}$,
with short-ranged coefficients $c_D$ and $c_E$ adjusted to $A\le 4$
nuclei. The ``1.8/2.0 (EM)'' interaction reproduces binding energies
and spectra in medium-mass and heavy
nuclei~\cite{hagen2016b,simonis2017,morris2018}, but yields too small
charge radii. The N$^2$LO$_{\rm sat}$ interaction yields radii in
better agreement with data.

\begin{figure}
	\begin{center}
		\includegraphics[width=\linewidth]{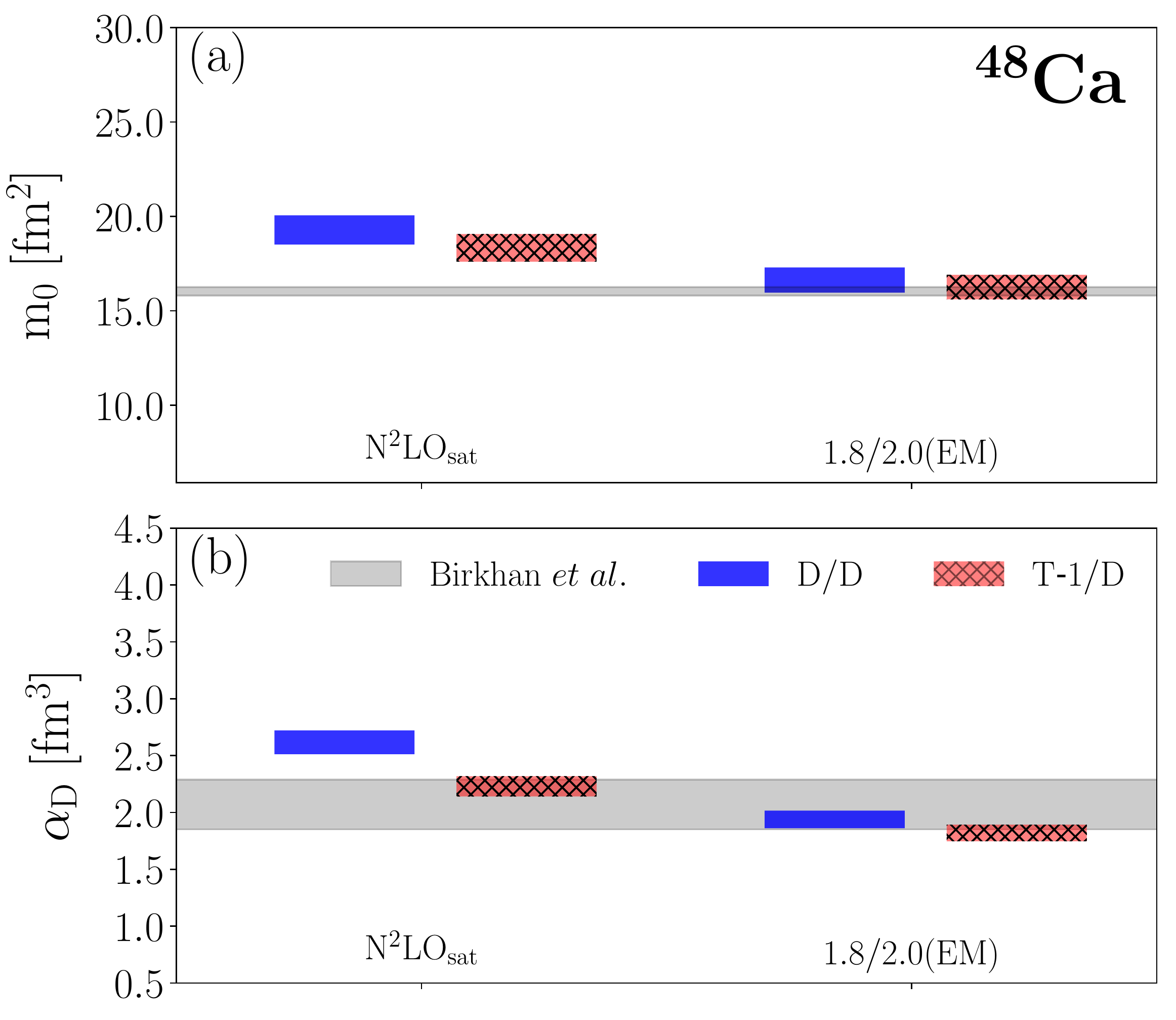}
    		\caption{(Color online) Comparison of $m_0$ (a) and
                  $\alpha_D$ (b) in the D/D (blue/left) and the
                  T-1/D (red/right) approximations in
                  $^{48}$Ca. Results for the N$^2$LO$_{\rm sat}$
                  (leftmost bands) and $1.8/2.0$ (EM) (rightmost
                  bands) interactions are shown. For the
                  polarizability the results are compared against
                  experimental data by \textcite{birkhan2017}.}
      \label{fig:48Ca_band}
  \end{center}
\end{figure}

Figure~\ref{fig:48Ca_band} shows the $m_0$ sum rule (top) and the
dipole polarizability $\alpha_D$ (bottom) obtained in the D/D and
T-1/D approaches using the N$^2$LO$_{\rm sat}$ (leftmost bands) and
$1.8/2.0$ (EM) (rightmost bands) interactions, respectively. The grey
horizontal bands are data. For N$^2$LO$_{\rm sat}$ we find that the
addition of $3p$-$3h$ excitations in the ground state using the T-1
approach improves the agreement with the data. As calculations of
excited states in the T-1 approach are computationally demanding and
currently not feasible at sufficiently large $E^F_{\rm 3max}$ cut, we
did not employ the T-1/T-1 approach for $^{48}$Ca. However, based on
our studies for $^4$He and $^{16}$O one may expect that the T-1/D and
T-1/T-1 approaches would yield similar results.

We observe that both for $m_0$ and $\alpha_D$ the T-1 approximation
for the ground state leads to a reduction of the strength, bringing
theory in agreement with the recent data from \textcite{birkhan2017}.
Remarkably, the inclusion of leading triples corrections also reduces
the Hamiltonian model dependence, leading to a more precise
calculation of $m_0$ and $\alpha_D$. The effect of the triples on
$\alpha_D$ amounts to 15$\%$ for the N$^2$LO$_{\rm sat}$ and 6$\%$ for
the $1.8/2.0$ (EM) interaction, which is consistent with the latter
being a much ``softer'' interaction. The triples effect on $m_0$ is
smaller, amounting to about 5$\%$ for N$^2$LO$_{\rm sat}$ and 2$\%$
for the $1.8/2.0$ (EM) interaction. We also performed T-1 calculations
for the charge radius of $^{48}$Ca, and found that inclusion of
triples excitations increases the radius by about 1\% for both the
$1.8/2.0$ (EM) and N$^2$LO$_{\rm sat}$ interactions as compared to the
D approximation. Thus, the effect of triples excitations on the charge
radius of $^{48}$Ca is neglible, and the results and conclusions of
Ref.~\cite{ekstrom2015} are accurate.

\begin{figure}[t]
  \begin{center}
    \includegraphics[width=1.0\linewidth]{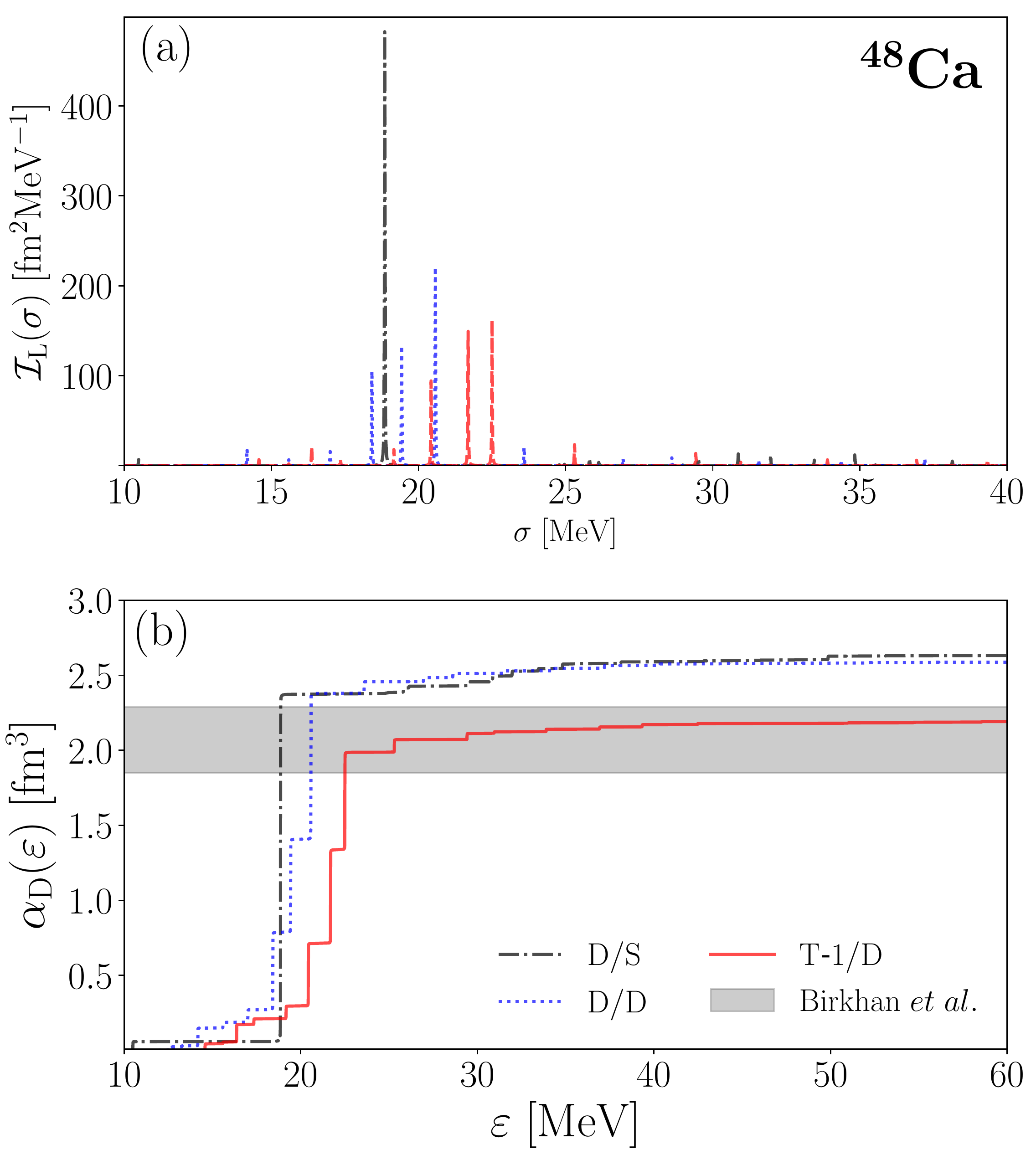}
    \caption{(Color online) Discretized response with $\Gamma=0.01$
      MeV (a) and running sum of $\alpha_D(\varepsilon)$ (b)
      for $^{48}$Ca with various coupled-cluster approximations
      compared with experimental results from \textcite{birkhan2017}}
      \label{fig:48Ca_running}
  \end{center}
\end{figure}

Figure~\ref{fig:48Ca_running} shows the Lorentz integral transform
(top) calculated for $\Gamma=0.01$ MeV using various coupled-cluster
approximations, and the running of the polarizability sum rule as a
function of $\varepsilon$ (bottom). The various approximation schemes
are shown in comparison with experimental data (gray bands). Besides
the D/D calculation, we present three different schemes with
increasing correlation order, namely D/S, T-1/D and T-1/T-1.

Interestingly, for $^{48}$Ca we see that for the D/S approximation the
strength is dominated by a single state at around 19~MeV of excitation
energy, while higher-order correlations included in D/D and T-1/D
shift and fragment the strength significantly. The inclusion of
$3p$-$3h$ excitations via the T-1 approach in the ground state, makes
the strength more fragmented and also increases some of the strength
at higher energy. Such fragmentation combined with the inverse energy
weight in the polarizability sum rule is the mechanisms that leads to
a reduction of $\alpha_D$. Opposite to what was found in $^{16}$O, for
$^{48}$Ca the computationally least expensive D/S approximation
coincidentally agrees with D/D and not with the computationally most
expensive approximation (T-1/D in this case).

\section{Conclusions}
\label{sec:conclusions}
We computed the electric dipole polarizability $\alpha_D$ of $^{48}$Ca
with an increased precision by including leading $3p$-$3h$
correlations.  Leading order $3p$-$3h$ excitations were implemented in
the CCSDT-1 approximation for the ground state, excited states, and
the similarity transformed dipole operator. The effect of $3p$-$3h$
excitations on the latter was found to be negligible.  The inclusion
of $3p$-$3h$ excitations in the ground state via CCSDT-1 is sufficient
to obtain an agreement with the hyperspherical harmonics approach for
$^4$He to better than 1\%. While the effect of triples is quite small
in $^4$He, it becomes larger in $^{16}$O and $^{48}$Ca. In $^{16}$O
triples excitations reduce the size of $m_0$ sum rule and $\alpha_D$
by about 4-6\%. In $^{48}$Ca they reduce $\alpha_D$ by about a 15\%
reduction for the N$^2$LO$_{\rm sat}$ interaction and by about 6\% for
the 1.8/2.0 (EM) interaction. This brings our coupled-cluster
calculations in agreement with the recent experimental result obtained
from inelastic proton scattering~\cite{birkhan2017}. The effect of
triples excitations is smaller for the $m_0$ sum rule in $^{48}$Ca,
and reduces it by about 5\% and 2\% for the two interactions
considered, respectively. Finally we found that the effect of triples
excitations on the charge radius of $^{48}$Ca is neglible, and
increases it by about 1\% for both interactions considered in this
work.

We note that some simpler approximations may occasionally coincide
with more sophisticated computations using the CCSDT-1 method both for
the ground and excited states. We conclude that the effect of
$3p$-$3h$ excitations in the ground state is more important than their
effect in excited states for the electromagnetic sum rules studied in
this work. The inclusion of triples excitations in the excited states,
while possible for light nuclei such as $^{4}$He and $^{16}$O, will
require further developments in order to overcome the hurdles
associated with the increase in computational cost for heavier nuclei.
We expect our results also to be relevant for other many-body
approaches, such as the self consistent Green's function and the in
medium similarity renormalization group methods.

\begin{acknowledgments}
  This work was supported in parts by the Natural Sciences and
  Engineering Research Council (NSERC), the National Research Council
  of Canada, by the Deutsche Forschungsgemeinschaft DFG through the
  Collaborative Research Center [The Low-Energy Frontier of the
    Standard Model (SFB 1044)], and through the Cluster of Excellence
  [Precision Physics, Fundamental Interactions and Structure of Matter
    (PRISMA)], by the Office of Nuclear Physics, U.S.~Department of
  Energy under Grants Nos.~DE-FG02-96ER40963 (University of Tennessee)
  DE-SC0008499 (SciDAC-3 NUCLEI), DE-SC0018223 (SciDAC-4 NUCLEI), and
  the Field Work Proposals ERKBP57 and ERKBP72 at Oak Ridge National
  Laboratory. Computer time was provided by the Innovative and Novel
  Computational Impact on Theory and Experiment (INCITE) program. This
  research used resources of the Oak Ridge Leadership Computing
  Facility located in the Oak Ridge National Laboratory, supported by
  the Office of Science of the U.S.~Department of Energy under
  Contract No.  DE-AC05-00OR22725, and computational resources of the
  National Center for Computational Sciences, the National Institute
  for Computational Sciences, and TRIUMF.
\end{acknowledgments}

\bibliography{master,refs_GH,refs}

\end{document}